\documentclass[12pt]{article}
\begin{document}
\title{From the Planck to the Photon Scale}
\author{B.G. Sidharth\\
International Institute for Applicable Mathematics \& Information Sciences\\
Hyderabad (India) \& Udine (Italy)\\
B.M. Birla Science Centre, Adarsh Nagar, Hyderabad - 500 063
(India)}
\date{}
\maketitle
\begin{abstract}
Using considerations from the Quantum Zero Point Field and
Thermodynamics, we show that the Planck Scale is the minimum
(maximum mass) and the Photon Scale is the maximum (minimum mass)
Scale in the universe. The arguments also deduce the residual cosmic
energy of $10^{-33}eV$ observed lately.
\end{abstract}
\section{Introduction}
It was argued by the author from different points of view that the
Photon would have a small mass $\sim 10^{-65}gms$ \cite{bhtd,mp}. We
will look into this now. This value is within the accepted
experimental limits for a Photon mass \cite{lakes}. It was further
argued that it is this Photon mass which is the source of the
puzzling residual cosmic energy that has been observed
lately\cite{mercini}.\\
Let us first derive this residual cosmic energy directly from the
background Dark Energy. We may reiterate that the "mysterious"
background Dark Energy is the same as the quantum Zero Point
Fluctuations in the background vacuum electromagnetic field  which
is described by harmonic oscillators \cite{bgsde}. Let us now
consider, following \index{Wheeler}Wheeler a \index{Harmonic
oscillator}Harmonic oscillator in its ground state remembering that
the background Zero Point Field is a collection of such oscillators
\cite{mwt}. The probability amplitude is
$$\psi (x) = \left(\frac{m\omega}{\pi \hbar}\right)^{1/4} e^{-(m\omega/2\hbar)x^2}$$
for displacement by the distance $x$ from its position of classical
equilibrium. So the oscillator fluctuates over an interval
$$\Delta x \sim (\hbar/m\omega)^{1/2}$$
The background \index{electromagnetic}electromagnetic field is an
infinite collection of independent oscillators, with amplitudes
$X_1,X_2$ etc. The probability for the various oscillators to have
amplitudes $X_1, X_2$ and so on is the product of individual
oscillator amplitudes:
$$\psi (X_1,X_2,\cdots ) = exp [-(X^2_1 + X^2_2 + \cdots)]$$
wherein there would be a suitable normalization factor. This
expression gives the probability amplitude $\psi$ for a
configuration $B (x,y,z)$ of the magnetic field that is described by
the Fourier coefficients $X_1,X_2,\cdots$ or directly in terms of
the magnetic field configuration itself by
$$\psi (B(x,y,z)) = P exp \left(-\int \int \frac{\bf{B}(x_1)\cdot \bf{B}(x_2)}{16\pi^3\hbar cr^2_{12}} d^3x_1 d^3x_2\right).$$
$P$ being a normalization factor. Let us consider a configuration
where the magnetic field is everywhere zero except in a region of
dimension $l$, where it is of the order of $\sim \Delta B$. The
probability amplitude for this configuration would be proportional
to
$$\exp [-(\Delta B)^2 l^4/\hbar c)$$
So the energy of \index{fluctuation}fluctuation in a region of
length $l$ is given by finally the density \cite{mwt,cr24,bgs}
\begin{equation}
B^2 \sim \frac{\hbar c}{l^4}\label{e1}
\end{equation}
The above energy density corresponds to an energy $\hbar c/l$ in the
volume $l^3$. This energy is minimum when $l$ is maximum. Let us
take $l$ to be the radius of the universe $\sim 10^{28}cms$. The
minimum energy residue of the background Dark Energy or Zero Point
Field now comes out to be $10^{-33}eV$, exactly the observed value.
This observed residual energy is a cosmic footprint of the
ubiquitous Dark Energy in the universe a puzzling footprint that, as
we noted, has recently been observed \cite{mercini}. If on the other
hand we take for $l$ the smallest possible length, which has been
taken to the Planck length $l_P$, as we will see in
the sequel, then we get the Planck mass $m_P$.\\
The minimum mass $\sim 10^{-33}eV$ or $10^{-65}gms$, will be seen to
be the mass of the Photon, which also is the minimum thermodynamic
mass in the universe, as shown by Landsberg from a totally different
point of view \cite{land}. So (\ref{e1}) gives two extreme masses,
the Planck mass and the Photon mass.\\
As an alternative derivation, it is interesting to derive a model
based on the theory of Phonons which are quanta of sound waves in a
macroscopic body \cite{huang}. Phonons are a mathematical analogue
of the quanta of the electromagnetic field, which are the Photons
that emerge when this field is expressed as a sum of Harmonic
oscillators. This situation is carried over to the theory of solids
which are made up of atoms that are arranged in a crystal lattice
and can be approximated by a sum of Harmonic oscillators
representing the normal modes of lattice oscillations. In this
theory, as is well known the Phonons have a maximum frequency
$\omega_m$ which is given by
\begin{equation}
\omega_m = c \left(\frac{6\pi^2}{v}\right)^{1/3}\label{e2}
\end{equation}
in (\ref{e2}) $c$ represents the velocity of sound in the specific
case of Photons, while $v = V/N$, where $V$ denotes the volume and
$N$ the number of atoms. In this model we write
$$l \equiv \left(\frac{4}{3} \pi v\right)^{1/3}$$
$l$ being the inter particle distance. Thus (\ref{e2}) now becomes
\begin{equation}
\omega_m = c/l\label{e3}
\end{equation}
Let us now liberate the above analysis from the immediate scenario
of atoms at lattice points and quantized sound waves due to the
Harmonic oscillations and look upon it as a general set of Harmonic
oscillators as above. Then we can see that (\ref{e3}) and (\ref{e1})
are identical as
$$\omega = \frac{mc^2}{\hbar}$$
So we again recover with suitable limits the extremes of the Planck
mass and the Photon mass (and other intermediate elementary particle
masses if we take $l$ as a typical Compton wavelength).\\
We now examine separately, the Planck scale and the photon mass. We
remark that there were basically two concepts of space which we had
inherited from the early days of modern science. The predominant
view has been the legacy from the Newtonian world view. Here we
consider space time to form a differentiable manifold. On the other
hand Liebniz had a different view of space, not as a container, but
rather made up of the contents itself. This lead to a view where
space time has the
smallest unit, and is therefore non differentiable.\\
Max Planck had noticed that, what we call the Planck scale today,
\begin{equation}
l_P = \left(\frac{\hbar G}{c^3}\right)^{\frac{1}{2}} \sim
10^{-33}cm\label{ea1}
\end{equation}
is made up of the fundamental constants of nature and so, he
suspected it played the role of a fundamental length. Indeed, modern
Quantum Gravity approaches have invoked (\ref{ea1}) in their quest
for a reconciliation of gravitation with other fundamental
interactions. In the process, the time honoured prescription of a
differentiable spacetime
has to be abandoned.\\
There is also another scale too, made up of fundamental constants of
nature, viz., the well known Compton scale,
\begin{equation}
l = e^2/m_ec^2 \sim 10^{-12}cm\label{ea2}
\end{equation}
where $e$ is the electron charge and $m_e$ the electron mass. This
had appeared in the Classical theory of the electron unlike the
Planck scale, which was
a product of Quantum Theory.\\
The scale (\ref{ea2}) has also played an important role in modern
physics, though it is not considered as fundamental as the Planck
scale. Nevertheless, the Compton scale (\ref{ea2}) is close to
reality in the sense of experiment, unlike (\ref{ea1}), which is
well beyond foreseeable direct experimental contact.
\section{The Planck and Compton Scales}
It is well known that String Theory, Loop Quantum Gravity and a few
other approaches start from the Planck scale. This is also the
starting point in the author's alternative theory of Planck
oscillators in the background dark energy. We first give a rationale
for the fact that the Planck scale would be a minimum scale in the
universe. Our starting point \cite{bgsijmpe} is the model for the
underpinning at the Planck scale for the universe. This is a
collection of $N$ Planck scale
oscillators.\\
Earlier, we had argued that a typical elementary particle like a
\index{pion}pion could be considered to be the result of $n \sim
10^{40}$ evanescent \index{Planck scale}Planck scale oscillators. We
will now consider the problem from a different point of view, which
not only reconfirms the above result, but also enables an elegant
extension to the case of the entire \index{Universe}Universe itself.
Let us consider an array of $N$ particles, spaced a distance $\Delta
x$ apart, which behave like oscillators, that is as if they were
connected by springs. We then have \cite{vandam,uof}
\begin{equation}
r  = \sqrt{N \Delta x^2}\label{De1d}
\end{equation}
\begin{equation}
ka^2 \equiv k \Delta x^2 = \frac{1}{2}  k_B T\label{De2d}
\end{equation}
where $k_B$ is the Boltzmann constant, $T$ the temperature, $r$ the
extent  and $k$ is the spring constant given by
\begin{equation}
\omega_0^2 = \frac{k}{m}\label{De3d}
\end{equation}
\begin{equation}
\omega = \left(\frac{k}{m}a^2\right)^{\frac{1}{2}} \frac{1}{r} =
\omega_0 \frac{a}{r}\label{De4d}
\end{equation}
We now identify the particles with \index{Planck}Planck
\index{mass}masses, set $\Delta x \equiv a = l_P$, the
\index{Planck}Planck length. It may be immediately observed that use
of (\ref{De3d}) and (\ref{De2d}) gives $k_B T \sim m_P c^2$, which
ofcourse agrees with the temperature of a \index{black hole}black
hole of \index{Planck}Planck \index{mass}mass. Indeed, Rosen had
shown that a \index{Planck}Planck \index{mass}mass particle at the
\index{Planck scale}Planck scale can be considered to be a
\index{Universe}Universe in itself. We also use the fact alluded to
that a typical elementary particle like the \index{pion}pion can be
considered to be the result of $n \sim 10^{40}$ \index{Planck}Planck
\index{mass}masses. Using this in (\ref{De1d}), we get $r \sim l$,
the \index{pion}pion \index{Compton wavelength}Compton wavelength as
required. Further, in this latter case, using (\ref{De1d}) and the
fact that $N = n \sim 10^{40}$, and (\ref{De2d}),i.e. $k_BT =
kl^2/N$ and  (\ref{De3d}) and (\ref{De4d}), we get for a
\index{pion}pion, remembering that $m^2_P/n = m^2,$
$$k_ B T = \frac{m^3 c^4 l^2}{\hbar^2} = mc^2,$$
which of course is the well known formula for the Hagedorn
temperature for \index{elementary particles}elementary particles
like \index{pion}pions. In other words, this confirms the conclusion
that we can treat an elementary particle as a series of some
$10^{40}$ \index{Planck}Planck \index{mass}mass oscillators. However
it must be observed from (\ref{De2d}) and (\ref{De3d}), that while
the \index{Planck}Planck \index{mass}mass gives the highest energy
state, an elementary particle like the \index{pion}pion is in the
lowest energy state. This explains why we encounter
\index{elementary particles}elementary particles, rather than
\index{Planck}Planck \index{mass}mass particles in nature. Infact as
already noted \cite{rr15}, a \index{Planck}Planck \index{mass}mass
particle decays via the \index{Bekenstein radiation}Bekenstein
radiation within a \index{Planck time}Planck time $\sim
10^{-42}secs$. On the other hand, the lifetime of an elementary
particle
would be very much higher.\\
In any case the efficacy of our above oscillator model can be seen by the fact that we recover correctly the \index{mass}masses and \index{Compton scale}Compton scales in the order of magnitude sense and also get the correct Bekenstein and Hagedorn formulas as seen above, and get the correct estimate of the \index{mass}mass of the \index{Universe}Universe itself, as will be seen below.\\
Using the fact that the \index{Universe}Universe consists of $N \sim
10^{80}$ \index{elementary particles}elementary particles like the
\index{pion}pions, the question is, can we think of the
\index{Universe}Universe as a collection of $n N \, \mbox{or}\,
10^{120}$ Planck \index{mass}mass oscillators? This is what we will
now show. Infact if we use equation (\ref{De1d}) with
$$\bar N \sim 10^{120},$$
we can see that the extent $r \sim 10^{28}cms$ which is of the order
of the diameter of the \index{Universe}Universe itself. Next using
(\ref{De4d}) we get
\begin{equation}
\hbar \omega_0^{(min)} \langle \frac{l_P}{10^{28}} \rangle^{-1}
\approx m_P c^2 \times 10^{60} \approx Mc^2\label{De5d}
\end{equation}
which gives the correct \index{mass}mass $M$, of the
\index{Universe}Universe which in contrast to the earlier
\index{pion}pion case, is the highest energy state while the
\index{Planck}Planck oscillators individually are this time the
lowest in this description. In other words the
\index{Universe}Universe itself can be considered to be described in
terms of normal modes of \index{Planck scale}Planck scale
oscillators (Cf.refs.\cite{psu,psp,uof,gip,ng} for details). We do
not need to specify $N$. We have in this case the following well
known relations
$$R = \sqrt{N}l, Kl^2 = kT,$$
\begin{equation}
\omega^2_{max} = \frac{K}{m} = \frac{kT}{ml^2}\label{ea3}
\end{equation}
In (\ref{ea3}), $R$ is of the order of the diameter of the universe,
$K$ is the analogue of
 spring constant, $T$ is the effective temperature while $l$ is the analogue of the
 Planck length, $m$ the analogue of the Planck mass and $\omega_{max}$ is the
 frequency--the
 reason for the subscript $max$ will be seen below. We do not yet give $l$ and $m$ their
 usual values as given in (\ref{ea1}) for example, but rather try to deduce these values.\\
We now use the well known result that the individual minimal
oscillators are black holes or mini universes as shown by Rosen
\cite{rosen}. So using the well known Beckenstein temperature
formula for these primordial black holes \cite{ruffini}, that is
$$kT = \frac{\hbar c^3}{8\pi Gm}$$
in (\ref{ea3}) we get,
\begin{equation}
Gm^2 \sim \hbar c\label{e4}
\end{equation}
which is another form of (\ref{ea1}). We can easily verify that
(\ref{e4}) leads to the value $m \sim 10^{-5}gms$. In deducing
(\ref{e4}) we have used the typical expressions for the frequency as
the inverse of the time - the Compton time in this case and
similarly the expression for the Compton length. However it must be
reiterated that no specific values
for $l$ or $m$ were considered in the deduction of (\ref{e4}).\\
We now make two interesting comments. Cercignani and co-workers have
shown \cite{cer1,cer2} that when the gravitational energy becomes of
the order of the electromagnetic energy in the case of the Zero
Point oscillators, that is
\begin{equation}
\frac{G\hbar^2 \omega^3}{c^5} \sim \hbar \omega\label{e5}
\end{equation}
then this defines a threshold frequency $\omega_{max}$ above in
which the oscillations become chaotic. In other words, for
meaningful physics we require that
$$\omega < \omega_{max}.$$
Secondly as we saw from the parallel but unrelated theory of phonons
\cite{huang,rief}, which are also bosonic oscillators, we deduce a
maximal frequency given by
\begin{equation}
\omega^2_{max} = \frac{c^2}{l^2}\label{e6}
\end{equation}
In (\ref{e6}) $c$ is, in the particular case of phonons, the
velocity of propagation, that is the velocity of sound, whereas in
our case this velocity is that of light. Frequencies greater than
$\omega_{max}$ in (\ref{e6}) are again meaningless.
We can easily verify that using (\ref{e5}) in (\ref{e6}) gives back (\ref{e4}).\\
Finally we can see from (\ref{e3}) that, given the value of $l_P$
and using the value of the radius of the universe, viz., $R \sim
10^{27}$, we can deduce that,
\begin{equation}
N \sim 10^{120}\label{e7}
\end{equation}
In a sense the relation (\ref{e4}) can be interpreted in a slightly
different vein as
representing the scale at which all energy- gravitational and electromagnetic becomes one.\\
It should also be noted that, a Planck scale particle is a
Schwarzchild Black Hole. From this point of view, we cannot
penetrate the Planck Scale - it constitutes a physical limit. Thus,
in this sense, the Planck scale is indeed the minimum scale while
the photon scale is the largest - that is, the concerned masses are
respectively the highest and lowest. 

\end{document}